\documentclass[preprint,showpacs,preprintnumbers,amsmath,amssymb]{revtex4}

\usepackage{graphicx}
\usepackage{dcolumn}
\usepackage{bm}

\begin{document}

\def\nn{\noindent}
\def\dd{\mathrm{d}}
\def\A{\mathrm{A}}
\def\B{\mathrm{B}}
\def\eps{\varepsilon}
\def\Erond{\mathcal{E}}
\def\LO{\text{LO}}
\def\SQL{\text{SQL}}
\def\rep{\mathrm{rep}}
\def\squeezing{\mathrm{squeezing}}
\def\sh{\mathrm{sh}}
\def\tof{\mathrm{tof}}
\def\ph{\mathrm{ph}}


\title{Quantum improvement of time transfer between remote clocks}

\author{Brahim Lamine}
 \email{brahim.lamine@spectro.jussieu.fr}
\author{Claude Fabre}%
\author{Nicolas Treps}
\affiliation{Laboratoire Kastler Brossel, Universit\'e Pierre et Marie
Curie-Paris 6,\\
ENS, CNRS; 4 place Jussieu, 75252 Paris, France}

\date{\today}

\begin{abstract}
Exchanging light pulses to perform accurate space-time positioning is a
paradigmatic issue of physics. It is ultimately limited by the quantum nature
of light, which introduces fluctuations in the optical measurements and leads
to the so-called Standard Quantum Limit
(SQL)~\cite{jaekel1996,lloyd2001,lloyd2004b}. We propose a new scheme combining
homodyne detection and mode-locked femtosecond lasers that lead to a new SQL in
time transfer, potentially reaching the yoctosecond range
($10^{-21}-10^{-24}\,$s). We prove that no other measurement strategy can lead
to better sensitivity with shot noise limited light. We then demonstrate that
this already very low SQL can be overcome using appropriately multimode
squeezed light. Benefitting from the large number of photons used in the
experiment and from the optimal choice of both the detection strategy and of
the quantum resource, the proposed scheme represents a significant potential
improvement in space-time positioning.
\end{abstract}

\maketitle

Accurate spacetime positioning has become a crucial issue for future space
experiments which require increasing resolution over large distances (see for
example~\cite{bender2002}). The position in space (by ranging to a reference)
or time (by clock synchronization with a reference) between two observers A and
B may be achieved through the Einstein protocol which consists to repeatedly
exchange light pulses~\cite{jaekel1996b}. From a fundamental point of view,
this procedure is at the root of Einstein's concept of space and time. From a
more practical point of view, it permits to distribute the time standard over
the whole earth and to precisely know the relative position of different
satellites in space.

\begin{figure}[htbp]
\includegraphics{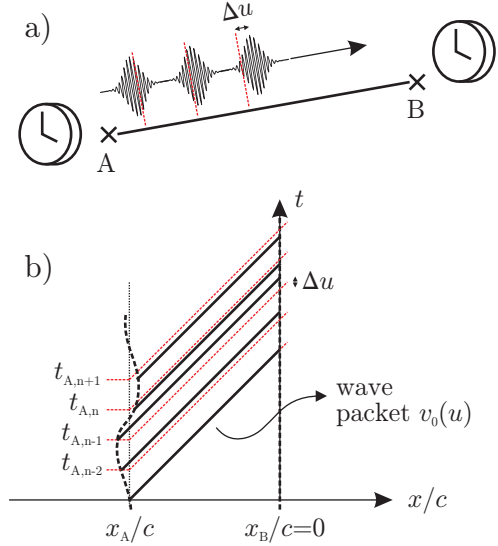} \caption{a) General
    scheme of a one way time transfer. We only consider propagation along the
    $x$ axis with diffraction neglected. b) Spacetime representation in
    the reference frame of observer B ($x_\B=0$). A modification
    $\Delta x_\A$ of the position of the observer A leads to a
    modification $\Delta u=-\Delta x_\A/c$ of the light cone variable
    that is emitted towards B and consequently leads to non regular time of
    arrival in B.}\label{fig:schéma général}
\end{figure}

The basic principle relies on the property that, in the absence of dispersion,
each pulse carries along its propagation a mean light cone variable $u=t\pm
x/c$ which remains constant so that the measurement of the time of arrival of
each pulse allows either a determination of distance or clock synchronization.
The generic situation considered in this paper is the following (see figure
(\ref{fig:schéma général}))~: observer A regularly emits light pulses at a rate
synchronized to its local clock; B receives these pulses and determine their
times of arrival by measuring the difference between the arrival times of the
incoming light pulses and light pulses delivered by a source located in B and
synchronized to a reference clock in B. The accuracy of this measurement relies
therefore on the precision of the clocks in A and B and on the sensitivity of
the determination of the delay between two light pulses, that we will show how
to optimize in the present paper.


Such a delay can be measured by at least two ways: the first one consists in
measuring the arrival time of the maximum of the pulse envelope. We will refer
to this procedure as a time-of-flight (tof) measurement. The second method
consists in using the information contained in the phase of the electric field
oscillation by making an interference pattern between the pulses arriving from
A and a Local Oscillator (LO) derived from the local clock in B. This pattern
will give the desired information if the phase of the pulse coming from A and
the phase of the LO in B are locked to their respective local clocks. This
method will be referred to as a phase (ph) measurement.

These measurement schemes suffer from quantum limits associated with the
quantum nature of light~\cite{jaekel1996}. For a coherent light pulse of
central frequency $\omega_0$ and frequency spread $\Delta \omega$, quantum
fluctuations lead to the so called Standard Quantum Limit (SQL) of ranging for
either time-of-flight~\cite{lloyd2001} or phase~\cite{caves1981,reynaud1990}
measurements. Those expressions are given by
\begin{equation}
(\Delta u)^\tof_\SQL=\frac{1}{2\Delta\omega\sqrt{N}}\quad,\quad (\Delta
u)^\ph_\SQL=\frac{1}{2\omega_0\sqrt{N}}\;.\label{SQL tof phase}
\end{equation}

\nn Where $N$ is the total number of photons measured in the experiment during
the detection time. Let us briefly discuss those two SQL. First, it is clear on
these expressions that the SQL can be as small as needed if one can use intense
enough light, but there are obvious practical limitations to the energy carried
by the light pulses. In contrast, isolated photons give rise to very low photon
fluxes, and the corresponding SQL is very quickly a limitation of experimental
protocols using photon counting techniques. The expressions also show that
optical frequencies lead to much smaller SQL than microwave frequencies because
of a larger $\omega_0$ and $\Delta\omega$. Finally, as $\omega_0>\Delta
\omega$, the phase method has a better ultimate sensitivity than the
time-of-flight technique but requires highly spatially and temporally coherent
sources.

For the time being, the resolution in time transfer is limited by classical
technical noises so that the previous SQL are not yet a limitation in time
transfer. Nevertheless, with the recent developments in stabilization of
frequency combs referenced to optical standard, it is getting closer and closer
to these quantum limits~\cite{ma2004}. Both for a fundamental point of view and
for future experiments, it is therefore necessary to compute the ultimate
sensitivity in time transfer with mode locked femtosecond laser since the
latter combine both a time-of-flight information in their enveloppe, and a well
stabilized phase information inside the enveloppe.

In order to compute the SQL in timing involving mode locked femtosecond lasers,
we begin by writing the positive frequency electric field operator $\hat
E^{(+)}_{(0)}$ emitted by A in the absence of any perturbations, as a
decomposition in temporal modes~:
\begin{equation}
\hat E^{(+)}_{(0)}(u)=\Erond\sum_n\hat a_n v_n(u)\quad,\quad\Erond=
i\sqrt{\frac{\hbar\omega_0}{2\eps_0 cT}}\;, \label{decomposition}
\end{equation}
\nn where $T$ is the measurement time. The orthonormal temporal modes $v_n(u)$
will be written as a (complex) time-varying amplitude $g_n(u)$ multiplied by a
propagation phase factor of the form $e^{-i\omega_0 u}$~:
\begin{equation}
v_n(u)= g_n(u)\,e^{-i\omega_0u}\;.
\end{equation}

The annihilation operator corresponding to those modes are noted $\hat a_n$.
Without any loss of generality, we can appropriately choose the mode basis such
that the mean value of the electric field operator $\hat E^{(+)}_{(0)}(u)$ is
proportional to $v_0$, namely $\langle\hat
E^{(+)}_{(0)}(u)\rangle=\Erond\sqrt{N}e^{i\theta}\,v_0(u)$, with $N$ the mean
number of photon and $\theta$ a global phase.

Now, any variation $\Delta u$ of the mean light cone variable, caused for
example by a distance change between A and B, leads to a modification of the
field received in B which reads $\hat{E}^{(+)}(u)=\hat{E}^{(+)}_{(0)}(u-\Delta
u)$ (see figure (\ref{fig:schéma général})). The temporal mode corresponding to
this field can be decomposed as follows if the perturbation $\Delta u$ is
small~:
\begin{equation}
v_0(u-\Delta u)\approx v_0(u)-\Delta u\left.\frac{\dd v_0(u)}{\dd
u}\right|_{u=0} = v_0(u)+\frac{\Delta u}{ u_0} w_1(u)\;. \label{mode perturbé}
\end{equation}

\nn The constant $u_0$ ensures the normalization of the new mode $w_1(u)$. The
latter one will be called the timing mode because it carries the timing signal
$\Delta u$. For pulses of frequency spread $\Delta\omega$~\footnote{We use the
standard definition of the statistical frequency
width~:$$\Delta\omega^2=\int_{-\infty}^{\infty}
\frac{\dd\omega}{2\pi}\,\omega^2|\tilde{g}_0[\omega]|^2=
\int_{-\infty}^{\infty}
\frac{\dd\omega}{2\pi}\,(\omega-\omega_0)^2|\tilde{v}_0[\omega]|^2$$\nn where
$\tilde{g}_0[\omega]$ is the Fourier transform of the enveloppe $g_0(u)$.},
$u_0$ is given by $u_0=1/\sqrt{\omega_0^2+\Delta\omega^2}$ and the expression
of the timing mode is
\begin{equation}
w_1(u)=\frac{1}{\sqrt{\alpha^2+1}}\big(i\alpha v_0(u)+v_1(u)\big)\qquad,\qquad
\alpha=\frac{\omega_0}{\Delta\omega}\;.\label{timing mode}
\end{equation}
$\alpha$ is roughly equal to the number of field oscillations within the pulse,
which can be as small as a few units for femtosecond pulses. The timing mode
$w_1(u)$ contains two terms: the first one, namely $iv_0(u)$, gives a
contribution to the timing signal via a phase change (interferometric method of
ranging). The second one, namely $v_1(u)$, is normalized and orthogonal to
$v_0$ so that it will be taken as the second mode of the basis $(v_n)_n$. It
reads~:
\begin{equation}
v_1(u)=-\frac{1}{\Delta\omega}\frac{\dd g_0(u)}{\dd u}\,e^{-i\omega_0
u}\;.\label{mode v1}
\end{equation}

\begin{figure}[htbp]
\includegraphics[width=8.6cm]{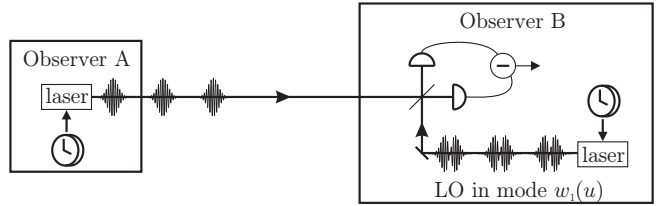} \caption{Proposed balanced homodyne scheme
to reach
    optimal detection in ranging measurement. The pulses synchronized
    on the clock in A are measured in B by homodyne detection with
    pulses synchronized on the local clock and in an adequate temporal mode
    (here is represented only the part $v_1$ of the LO for clarity).}
  \label{fig:homodyne detection}
\end{figure}

This mode gives a contribution to the timing signal via a time shift of the
pulse enveloppe (time-of-flight technique). The latter mode is represented in
the figure (\ref{fig:homodyne detection}) and is the temporal analog of the
spatial TEM$_{01}$ gaussian mode when the emitted pulses are gaussian.


The timing signal $\Delta u$ can be retrieved by projecting $v_0(u-\Delta u)$
on the timing mode $w_1(u)$. This can be done using the balanced homodyne
detection scheme represented in figure (\ref{fig:homodyne detection}) where the
input pulses are mixed with a Local Oscillator (LO) put in the timing mode
$w_1$~\cite{delaubert2006}, so that
$\langle\hat{E}^{+}_\LO(u)\rangle=\Erond\sqrt{N_\LO}e^{i\theta_\LO}w_1(u)$,
with $N_\LO$ the mean number of photon in the LO field and $\theta_\LO$ its
phase. Denoting $(\hat{b}_n)_n$ the annihilation operators for the LO, the
homodyne signal $\hat{D}$ reads~:
\begin{equation}
\hat{D}=|\Erond|^2\sum_n\big(\hat a_n^\dagger\hat{b}_n+\hat{b}_n^\dagger\hat
a_n\big)\;.
\end{equation}

The mean signal of the balanced homodyne detection when a timing offset $\Delta
u$ is present, then reads~:
\begin{equation}
\langle\hat D\rangle=2|\Erond|^2\sqrt{NN_\LO}\left[\frac{\Delta u}{u_0}
\,\cos(\theta-\theta_\LO) +\frac{\alpha}{\sqrt{\alpha^2+1}}
\,\sin(\theta-\theta_\LO)\right]\;.\label{ranging signal}
\end{equation}

We assume from now on that, as usual, the LO is much more intense than the
input field. The general case can be treated without difficulty. In this
situation, the variance of the balanced homodyne signal, taken for $\Delta
u=0$, is given by~:
\begin{equation}
\sigma^2_{\hat{D}}\equiv\langle\delta\hat
D^2\rangle=\frac{|\Erond|^4N_\LO}{1+\alpha^2}\left(\alpha^2\sigma^2_{\hat{P}_0}+
\sigma^2_{\hat{Q}_1}\right)\;,\label{variance}
\end{equation}
\nn where $\sigma^2_{\hat{P}_0}$ and $\sigma^2_{\hat{Q}_1}$ are the variances
of the quadrature operators $\hat{P}_0$ (phase operator of mode $v_0$) and
$\hat{Q}_1$ (amplitude operator of mode $v_1$) of the input field~:
\begin{equation}
\hat{P}_0=i\left(\hat{a}^\dagger_0e^{i\theta_\LO}-\hat{a}_0e^{-i\theta_\LO}
\right) \qquad\text{and}\qquad \hat{Q}_1=\hat{a}^\dagger_1e^{i\theta_\LO}+
\hat{a}_1e^{-i\theta_\LO} \;.
\end{equation}

The Standard Quantum Limit (SQL) is then obtained as the smallest $\Delta u$
that can be measured using shot noise limited coherent light
($\sigma^2_{\hat{P}_0}=\sigma^2_{\hat{Q}_1}=1$), assuming a signal to noise
ratio equal to one ($\langle\hat D\rangle=\sigma_{\hat{D}}$). It is obtained
for $\theta=\theta_\LO$ and is given by~:
\begin{equation}
(\Delta u)_{\SQL}=\frac{1}{2\sqrt{N}
\sqrt{\omega_0^2+\Delta\omega^2}}\;.\label{SQL}
\end{equation}
The expression (\ref{SQL}) is one the main results of this paper and gives a
new SQL in timing. The latter is lower than both the SQL in time-of-flight and
phase measurements (see equation (\ref{SQL tof phase})), which obviously are
special cases of our scheme when the LO is either in the $iv_0$ or $v_1$ mode.
This means that the proposed balanced homodyne detection scheme has a better
sensitivity than existing schemes based on either time-of-flight or
interferometric measurement. The improvement comes from the fact that coherent
pulses, in addition to their phase, carries a time of flight information in
their time varying enveloppe. Both informations are read by the balanced
homodyne detection if the LO is shaped in the mode $w_1(u)$. Let us stress that
such optimized measurements have already been successfully employed for pure
phase measurement~\cite{berry2004} and in the spatial domain to measure
transverse beam displacement and tilt~\cite{treps2005}.

For a $P=10\,$mW laser with $\lambda\simeq810\,$nm and a $10\,$fs pulse
duration, the SQL is equal to $(\Delta u)_\SQL=2\times 10^{-23}\,s$, i.e. a
noise level of $2\times 10^{-23}\,s/\sqrt{\text{Hz}}$ (20 yoctoseconds for one
second integration time).

A natural question is to know whether it is possible to reach still better
sensitivity on the same beam but by using another measurement strategy. An
answer can be provided in the context of information theory with the help of
the Cramer-Rao bound~\cite{Refregier:book}, which gives the smallest measurable
delay $\Delta u$ that can be achieved in the presence of a given distribution
of noise. This bound has the property of being independent of the measurement
strategy and depends only on the noise of the incoming signal. A calculation of
the Cramer-Rao bound, analogous to the one detailed
in~\cite{delaubert:PhD,delaubert2008} proves that using coherent light, this
bound is precisely equal to the expression (\ref{SQL}) of $(\Delta u)_{\SQL}$.
We are therefore sure that no other measurement scheme will reach a better
accuracy than the introduced balanced homodyne detection and in this sense this
scheme is said to be efficient.

Obviously the SQL (\ref{SQL}) is the fundamental limit when one restricts
oneself to the use of classical states of light and coherent states, as proven
with the previous standard Cramer-Rao bound. Nevertheless, it is well known
that it can be beaten using quantum
resources~\cite{lloyd2004b,dowling2000,bartlett2005}. For example, the
improvement of the sensitivity in interferometric measurements using squeezed
light has been proposed~\cite{caves1981,reynaud1990}, observed
experimentally~\cite{kimble1987,grangier1987,barnett2003}, and will be
certainly practically implemented in the future generations of interferometric
detectors of gravitational waves~\cite{mckenzie2002}. The use of an entangled
photon source to improve time-of-flight ranging measurements in the
photon-counting regime has been also proposed~\cite{lloyd2001,lloyd2002a} and
experimentally demonstrated~\cite{valencia2004} at a picosecond level of timing
sensitivity. We propose here to improve the scheme introduced previously by
using appropriately squeezed light.

Inspection of equation(\ref{variance}) immediately shows that in the case of a
strong LO the signal to noise ratio is increased if the noise of the incoming
mode $w_1$ is below the shot noise. This can be obtained if squeezing of the
input field modes $v_0$ and $v_1$ is achieved along the quadratures $\hat{P}_0$
and $\hat{Q}_1$ respectively. This therefore requires to first squeeze the
phase of the input field and mix it with a squeezed vacuum mode
$v_1$~\cite{valcarcel2006}, using procedures already demonstrated in the
spatial domain~\cite{delaubert2006}. If we assume that the squeezing
coefficient is equal for the two states, namely
$\sigma_{\hat{X}_0}=\sigma_{\hat{X}_1}=e^{-r}$ ($r\geq1$ being the squeezing
parameter), then the new minimum measurable value of $\Delta u$ is given by~:
\begin{equation}
(\Delta u)_\squeezing=\frac{1}{2\sqrt{N}\sqrt{\omega_0^2+\Delta\omega^2}}
\,e^{-r}\;.\label{SQL squeezing}
\end{equation}

This minimum resolvable $\Delta u$ is thus reduced below the SQL (\ref{SQL}) by
the factor $e^r$. Note that the expression for the general case of different
squeezing along $\hat{X}_0$ and $\hat{X}_1$, as well as a LO not supposed
strong, can be obtained straightforwardly from the equations given in the
paper.

Using the best present technology, the noise reduction factor can reach
$10\,$dB~\cite{vahlbruch2007,takeno2007}, i.e. a factor of $10$ improvement,
even at low noise frequencies. The advantage of squeezing over the other
proposed quantum techniques such as entanglement is that it can be used
together with an intense beam for which the SQL is already very low. In
addition the squeezed beam travels along with the signal beam and therefore
they both share commun noises. This is not the case for protocol using non
local quantum correlations (such as most of the protocols based on
entanglement) where the quantum correlations are spread over a large region of
space and submitted to differential noise effects. The main drawback of
squeezing is its sensitivity to losses in the optical system and the detectors.
This means that the technique could be used in situations where light
propagates in vacuum, for example between satellites in flying formation.

An experimental implementation of the scheme with the aim at reaching the SQL
and then observe the quantum improvement suffers different technical
challenges. First of all, reaching a timing precision in the yoctosecond regime
requires very stable laser repetition rate and phase stabilization. This can be
eventually be achieved with mode-locked femtosecond lasers which are already
used for absolute and relative ranging in different measurement
schemes~\cite{ye2004,minoshima2000,towers2004,joo2004}. The dominant source of
noise in equation (\ref{variance}) is given by the noise $\sigma_{\hat{X}_0}$
of the phase of $v_0$. Self-referencing stabilization using a $f-2f$ beat
allows to keep this noise to a very low level, down to
$10^{-5}\,$rad/$\sqrt{\mathrm{Hz}}$ at $10^5\,$Hz with state-of-the-art
stabilization techniques~\cite{fortier2002,bartels2004}, corresponding to a
timing noise of $4\times10^{-21}\,$s/$\sqrt{\mathrm{Hz}}$ at $10^5\,$Hz.
Concerning the repetition rate $T_\rep$, the latter can be locked to an optical
reference, and current technology leads to a time jitter noise level of
$10^{-18}\,$s/$\sqrt{\mathrm{Hz}}$ at
$10^{5}\,$Hz~\cite{bartels2003,shelton2002,schibli2003}. Another experimental
challenge is to produce the squeezed temporal mode $w_1$. Indeed, if the mode
$w_1$ can be obtained with presently available commercial mode
shapers~\cite{sato2002}, the squeezing is much more challenging, but can in
principle be obtained by propagation through a non-linear Kerr
medium~\cite{schmitt1998} or more efficiently by using parametric down
conversion pumped by mode-locked lasers~\cite{slusher1987,rosenbluh1991}, or
even synchronously pumped OPOs~\cite{valcarcel2006}.

\begin{acknowledgments}
We are grateful to Serge Reynaud and Vincent Delaubert for fruitful
discussions. Laboratoire Kastler Brossel is unit\'e mixte de recherche (UMR)
n$^\circ 8552$ of the CNRS.
\end{acknowledgments}

\bibliography{TimeTransferPRL}

\end{document}